\documentclass{article}

\usepackage{caption}
\usepackage{arxiv}
\usepackage{graphicx}
\usepackage[utf8]{inputenc} % allow utf-8 input
\usepackage[T1]{fontenc}    % use 8-bit T1 fonts
\usepackage{hyperref}       % hyperlinks
\usepackage{url}            % simple URL typesetting
\usepackage{booktabs}       % professional-quality tables
\usepackage{amsfonts}       % blackboard math symbols
\usepackage{nicefrac}       % compact symbols for 1/2, etc.
\usepackage{microtype}      % microtypography
\usepackage{lipsum}
\usepackage{multicol}
\usepackage{multirow}
\usepackage{subcaption}
\usepackage{arxiv}
\usepackage[square, numbers, comma, sort&compress]{natbib}

\title{Sequential image processing methods for improving semantic video segmentation algorithms}

\author{
  Beril Sirmacek, Nicol\`{o} Botteghi, Santiago Sanchez Escalonilla Plaza \\
  Robotics and Mechatronics\\
  University of Twente\\
  Enschede, The Netherlands \\
  \texttt{(b.sirmacek, n.botteghi, s.sanchezescalonillaplaza)@utwente.nl} \\
}

\begin{document}
\maketitle

\begin{abstract}
Recently, semantic video segmentation gained high attention especially for supporting autonomous driving systems. Deep learning methods made it possible to implement real-time segmentation and object identification algorithms on videos. However, most of the available approaches process each video frame independently disregarding their sequential relation in time. Therefore their results suddenly miss some of the object segments in some of the frames even if they were detected properly in the earlier frames. Herein we propose two sequential probabilistic video frame analysis approaches to improve the segmentation performance of the existing algorithms. Our experiments show that using the information of the past frames we increase the performance and consistency of the state of the art algorithms.
\end{abstract}

% keywords can be removed

\section{Introduction}
In an era for automation, it is not far fetched to think of a scenario where transportation does not suppose a hustle for the driver anymore. Regarding to commuting statistics, it is interesting to take a look at the American panorama as the living patterns are more standardized than over the different countries in Europe. A recent study by Statista \citep{drivingstats} states that in America in 2016, an estimated 85.4 percent of 150M workers drove to their workplace in an automobile, while only 5.1 percent used public transportation for this purpose. Out of the 85.4 percent, a total of 77 percent (115M people) drove alone to work everyday \citep{drivingstats2}.\par
Although some people enjoy the act of driving, it is fair to generalize that driving during rush hour is considered to be one of the most stressful scenarios for the daily commuting. While passengers can just sit and relax, the driver has to be constantly conscious about his actions during the whole ride. Self driving cars aim to free the driver from this activity, allowing him to spend his time in more valuable tasks. AC can also transform the current traffic system scenario by making it safer and more efficient to navigate, extending the benefits of automation to non-AC users.\par
Autonomous navigation however, is not a recent invention. In 1912, Lawrence Sperry successfully demonstrated the implementation of an autopilot-system on aviation. In an aircraft exhibition celebrated in Paris in 1914, Sperry performed numerous in-flight tricks in front of an audience to test the autonomy of the navigation system under no pilot conditions.\par
However, solving Autonomous navigation problems for cars, drones, bus, trucks... is not a trivial problem for different reasons:
\begin{itemize}
    \item From the structural point of view, to list some examples: non-standardize roads (undefined or different lane sizes), inconsistent driving conditions (changes in weather, driving surface might deteriorate), obstacles or debris, ambiguous drivable space, undefined traffic signs location.
    \item From the non-structural point of view other factors come into play, such as: human or animal interaction (unpredictable behavior).
\end{itemize}
The eruption of deep learning on the last decade has allowed to create safer and more intelligent pilot systems that enable autonomous vehicles to operate better than under previously unseen scenarios. Deep learning together with the motivation of some companies to take autonomous navigation systems into mass production makes the present time to be the perfect one to solve the autonomous vehicles enigma.

\subsection{Autonomous Navigation}
Although any system that requires autonomous navigation (cars, drones or any other mobile robot) can be considered for this topic, this document will focus on autonomous cars. The reason for this focus is that the late outburst of autonomous navigation in the automobile industry is promoting the scientific interest towards autonomous cars resulting in new studies and data sets that cover this application.\par
When talking about autonomy, the National Highway Traffic Safety Administration (NHTSA) has defined the following levels of car automation:
\begin{itemize}
    \item{Level 0:} No Automation. The driver performs all driving tasks.
    \item{Level 1:} Driver Assistance. The Vehicle is controlled by the driver, but some driving assists features may be included in the vehicle design (such as ESP, Airbags, Lane keeping,...)
    \item{Level 2:} Partial Automation. Driver-assist systems that control both steering and acceleration/deceleration, but the driver must remain engaged at all times (e.g. cruise control or parking assistance).
    \item{Level 3:} Conditional Automation. The driver is a necessity but not required at all times. He must the ready to take control of the vehicle at all times with notice.
    \item{Level 4:} High Automation. The vehicle is capable of performing all driving functions under certain conditions. The driver may have the option to control the vehicle.
    \item{Level 5:} Full Automation. The vehicle is capable of performing all driving functions under all conditions. The driver may have the option to control the vehicle.
\end{itemize}
There are different companies trying to adapt classic vehicles to the different levels of automation. Nowadays most of the cars available have at least a level 2 of automation making level 3 the next step of the challenge.\par 
Level 3 is currently dominated by Tesla, since the release of autopilot in 2016, Tesla has been manufacturing new vehicles surpassing the 1 Billion miles mark driven autonomously (followed by Waymo with 10 Million miles). Despite this big improvement, further levels of automation require a deeper study of the current technology and gather big amounts of data from driving patterns and uncommon situations. \par
In order to grant cars with autonomy, former cars need to be upgraded both in the hardware as well as in the software side. A key piece for this upgrade is the choice of the car's equipment. Cameras are the most common sensor present in autonomous vehicles, which along with other type of sensors are able to recreate a virtual representation of the surroundings.

\begin{figure}[ht]
    \centering
    \includegraphics[scale=0.3]{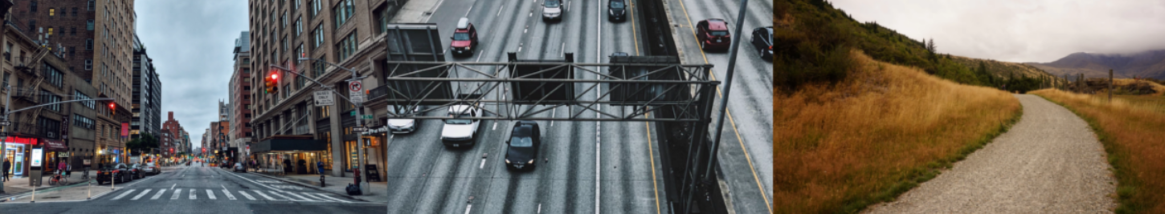}
    \caption{Different types of roads. From left to right: urban road, highway and rural road.}
    \label{fig:Landscapes}
\end{figure}

% \subsubsection{Driving domain}
The application domain of autonomous driving extends to anyplace with a drivable area (figure \ref{fig:Landscapes}). Apart from the variety of roads, the difficulty of automation is enhanced by the bounds of the problem: outdoors application. This loose definition of the domain specifications is what makes autonomous driving so challenging. \par

Granting a machine with the capacity of overtaking humans for tasks such as transportation is a non-trivial problem. Driving is a life-risk activity and therefore needs of a meticulous study, testing and evaluation of these new autonomous technologies.

\subsection{Semantic image segmentation}
Computer Vision is the field of engineering that focuses on the extraction of the information encoded on images for its use on different applications. Classical computer vision extracts this information through the calculation of different image descriptors. The calculation of the image descriptors is conditioned by the system characteristics: image resolution, object shape, light conditions, application domain. The process that defines the image descriptors derivation is called feature extraction.\par
Image descriptors are usually designed as hand-engineered filters, providing solutions that are rather rigid (application specific) and reliable only under very restricted conditions. Unfortunately autonomous navigation falls into a completely opposite scenario, requiring of applications that can perform robustly under very dynamic circumstances.\par
The main advantage of deep learning is its flexibility to generalize to previously unseen data. Since the application of Convolutional Neural Networks (CNNs) \citep{LeCun1999ObjectRW} for image processing, Deep Learning has been the protagonist on countless Computer Vision conferences and research papers. CNNs allow the extraction of features in a more efficient and meaningful way than classical image descriptors, based on image gradient calculations. Standing out due to their capacity of automation of image descriptors, CNNs allow the creation of Image Processing applications with a high level of abstraction and accuracy.\par
\begin{figure}[ht]
    \centering
    \includegraphics[scale=0.3]{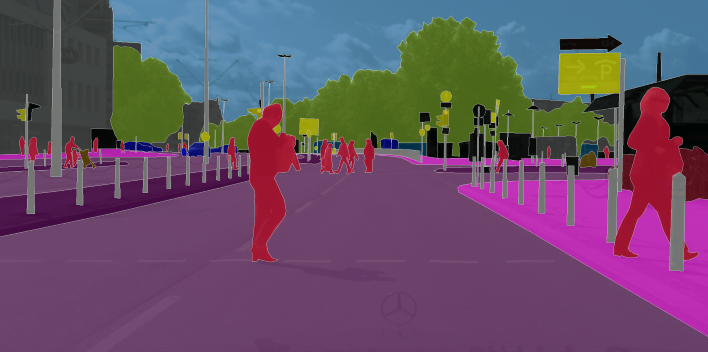}
    \caption{Ideal result of a semantic image segmentation. In this figure, all the objects that conform the image are perfectly classified into the different colors that define each category: road, sidewalk, pedestrian, tree, building or traffic sign. This image is part of the Cityscapes groundtruth densely annotated dataset. Cityscapes is a large-scale data set created as tool to evaluate the performance of vision algorithms intended for semantic urban scene understanding and help researchers to exploit large volumes of annotated data. Image source: \cite{cityscapesweb}.}
    \label{fig:cityscapesgt}
\end{figure}

Semantic image segmentation (figure \ref{fig:cityscapesgt}) is just one of the many Deep Neural Networks (DNN) applications. The goal Semantic Segmentation application is to detect and classify objects in the image frame by applying pixel-level classification of an input image into a set of predefined categories. Semantic image segmentation provides a level of scene understanding much richer than any other detection algorithms, it includes detailed information about the shape of the object and its orientation. Semantic segmentation can be used in autonomous navigation to precisely define the road (or drivable space) and its conditions (erosion, presence of obstacles or debris), it is also very useful for navigation in crowded areas being able to accurately calculate the gap between obstacles and even make predictions of the future position of the obstacles based on its shape and trajectory. Semantic image segmentation models can generally be divided into two parts: the feature extraction layers (hidden layers) and the output layers. The feature extraction layers use CNNs along with other techniques such as pooling or skip connections to obtain a low level representation of the image. And the output layers create the necessary relations to draw-out the pixel classification.\par
The scope of the project will be restricted to the analysis of a hypothetical video feed coming from the frontal camera of an autonomous car. The purpose of this camera is to elaborate a frontal representation of the environment that can be used for navigation. Flying objects, dirt or sun glare are some of the external factors that can affect the correct performance of cameras. In order to guarantee the passengers' safety, the detection system of autonomous cars must stand out for the robustness and consistency of its results and all these situations need to be considered. An additional observation is that when applied to autonomous navigation applications, the segmentation should prevail the detection of obstacles over driving space to ensures the avoidance collisions.\par

\section{Problem statement} \label{sec:problemStatement}
Semantic segmentation allows autonomous cars to obtain an accurate representation of the outside world. This representation is used to define the available navigation space and the presence of obstacles necessary to calculate navigation trajectories.\par
\begin{figure}[ht]
    \centering
    \includegraphics[scale=0.35]{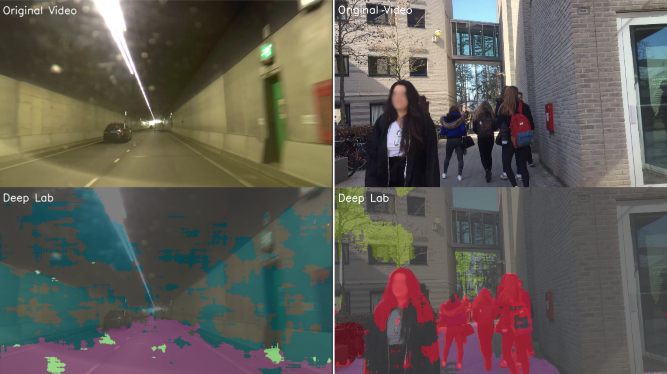}
    \caption{Figure illustration of the segmentation level obtained by DeepLabv3 \citep{Chen2017RethinkingAC}, the current state-of-the-art semantic image segmentation model trained on the Cityscapes dataset \citep{Cordts2016Cityscapes}. This figure illustrates two different levels of segmentation imperfection. Left image: example of a totally missed classification of the cars in front of the camera, added to a noisy classification of the road and the walls. Right image: example of a partial segmentation of the pedestrians.}
    \label{fig:deeplab-baseline}
\end{figure}

Figure \ref{fig:cityscapesgt} shows an example of a perfect semantic image segmentation, however it is very difficult to obtain a segmentation in such a high level of detail. The deep learning model would require large amounts of high resolution finely annotated and varied data to allow the training optimization algorithm reach the desired accuracy while not overfitting. In contrast, figure \ref{fig:deeplab-baseline} shows a real example of how an image that has been processed using an out-of-the-box state-of-the-art semantic image segmentation model (DeepLabv3 \citep{Chen2017RethinkingAC}) that was trained on the Cityscapes data set \citep{Cordts2016Cityscapes} looks like.  
Figure \ref{fig:deeplab-baseline} illustrates how the current level of a semantic image segmentation state-of-the-art model performance differs from the ground-truth example shown on figure \ref{fig:cityscapesgt}. Although a partial classification might be good enough for obstacle avoidance, in some cases the semantic image segmentation model completely misses the classification of the obstacle and therefore can cause an accident. For this reason, autonomous cars have numerous sensors that allow the detection of obstacles at different distance ranges not relying on bare-semantic image segmentation models as the main source of information.\par
Another of the effects that can be observed after applying semantic image segmentation models for the analysis of videos is temporal inconsistency. Analyzing a video frame by frame causes a segmentation that is not consistent over time, small variations in the frame produce high variances in the segmentation .\par
This study examines \textit{\textbf{how to reduce the incorrect classifications produced by semantic image segmentation models by combining the information of neighbouring frames}}
 In an attempt to improve the obstacle detection, this study can be broken down into the following research questions: 
\begin{itemize}
    \item \emph{Analysis of the state-of-the-art: \textbf{what is the current state-of-the-art for semantic image segmentation?}}
    \item \emph{Temporal extension: \textbf{how to extend semantic-image-segmentation models for the analysis of sequences?}}
    \item \emph{Reducing missed classifications: \textbf{what kind of mechanisms can be applied to reduce the number of false classifications?}}
\end{itemize}

\section{Background}

As previously stated, one of the main goals of this article is \textit{\textbf{how to extend semantic image segmentation models for the analysis of sequences}} (videos are a sequence of images). The main difference between images and videos is that the latter consists of a group of images (frames) that are adjacent in time, indirectly encoding a temporal history. In order to exploit the sequential information present in videos, this section will introduce the available tools capable of modeling sequences. In section \ref{ch:four}, some of these techniques will be used with a semantic image segmentation model in an attempt to add the video temporal information into the segmentation.\par
Given a causal system, sequence modeling consists on elaborating a model that is able to reproduce the dynamic behavior present in the observed data. From probabilistic methods to neural networks, this section summarizes different procedures used to capture temporal dynamics.\par
The methods reviewed in this section can be divided into two different groups depending on the tools used for sequence modeling: Conditional Probability and Deep Learning Architectures. The first one reviews causal modeling using probability relations. The second one introduces deep neural networks that have been specifically designed for modeling videos. 

\subsection{Sequence Modeling: Conditional probability}\label{sec:conditionalprob}
This section analyzes how to model the association between variables using probabilistic relations. Given two observed variables '$x_1$' and '$x_2$', the conditional probability of '$x_2$' taking a value given that '$x_1$' takes another value (two different events) is is defined as \citep{Murphy2012MachineL}:
\begin{equation}
P(x_2|x_1) = \frac{P(x_1, x_2)}{P(x_1)}\textit{, if $P(x_1)>0$}
\label{eq:x2givenx1}
\end{equation}
Where the numerator of equation \ref{eq:x2givenx1} is the joint probability of both events happening at the same time, and the numerator is the marginal probability of event '$x_1$. It is possible to extend this notation to cover a bigger set of events (or a sequence of events). For a set of variables $X_t = x_1, x_2,... x_t$ (for $t>0$), the probability of the variable $x_t$ conditioned to the rest of the variables in '$X_t$' is:
\begin{equation}
P(x_t|\{X_{\tau}, \tau \neq t\}) = \frac{P(x_1,x_t)}{P(x_1,...x_{(t-1)})}=\frac{P(X_t)}{P(\{X_{\tau}, \tau \neq t\})}\textit{, if $P(\{X_{\tau}, \tau \neq t\})>0$}
\label{eq:xtgivenxt-1}
\end{equation}
Besides expressing the relation between variables using conditional probability notation, it is also possible to use graphical models or probabilistic graphical models. A Probabilistic Graphical Model (PGM) is a graph that expresses the conditional dependence relation between random variables. Conditional probability notation in combination with probabilistic graphical models are commonly used in fields such as probability theory, statistics and machine learning.\par
\begin{figure}[ht]
    \centering
    \includegraphics[scale=0.35]{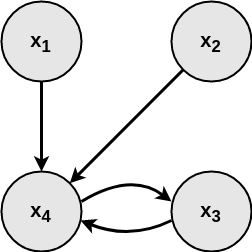}
    \caption{Possible probabilistic graphical model of equation \ref{eq:xtgivenxt-1} for $t=4$. In this graph, '$x_4$' depends on '$x_1$, $x_2$ and $x_3$'; '$x_3$' depends on '$x_4$'; and '$x_1$ and $x_2$' are independent.}
    \label{fig:PGM}
\end{figure}

A possible graphical representation of equation \ref{eq:xtgivenxt-1} for $t=4$ can be found in figure \ref{fig:PGM}
There are two main approaches that can be followed when defining the probability model for a dynamic system: a generative approach or a discriminative approach. Although the final purpose of both approaches is the same, to sample data from a probability distribution, each approach is different.\par
Generative models focus on modeling how the data is generated, in other words, modelling the joint probability $P(X_t)$, where '$X_t$' is the set of variables involved in the process-- e.g. in retrospect to the analysis of videos, each variable in '$X_t$' can represent the value of a pixel over consecutive time steps, and '$t$' the frame index. A generative model is able to calculate the joint distribution of all the present variables '$P(X_t)$'. For the simple case of having 4 variables, modeling '$P(X_4)$' allows finding all the possible combinations for these variables: having observed the pixel at times 1, 2 and 3, it is possible to estimate $x_4$; or any other combination, such as computing $x_2$ given $x_1$, $x_3$ and $x_4$.\par
On the other hand discriminative models only focus on modeling the conditional relation between variables (equation \ref{eq:xtgivenxt-1}), not paying attention on how the data is generated-- e.g. having observed $x_1$, $x_2$ and $x_3$ it is possible to estimate $x_4$ but it does not allow to compute any other combination for the variables.\par

% Generative models can produce synthetic data
From the definition, generative models may appear to be more general and insightful about the present dynamic system than the discriminative ones. However, discriminative models are often preferred for classification tasks such as logistic regression \citep{generativeVSdiscriminative}. The reason for this preference is that the generalization performance of generative models is often found to be poorer than the one proper of discriminative models due to differences between the model and the true distribution of data \citep{generativeVSdiscriminative2}.\par
The most common procedures to create probability models will be reviewed in the following order: Naive Bayes Classifiers, Markov Chains and Hidden Markov Models (HMM).
% and Conditional Random Fields (CRF).

\subsubsection{Naive Bayes Classifier}\label{sec:bayes}
The Naive Bayes classifier is a generative approach because it models the joint probability '$P(X_t)$' and afterwards calculates the conditional probability applying the Bayes Rule. Starting from the definition of conditional probability (equation \ref{eq:xtgivenxt-1}), it is possible to apply the product rule of probability to the numerator, '$P(X_t)$' as:
\begin{equation}
P(X_t) = P(\{X_{\tau}, \tau \neq t\}, x_t) = P(\{X_{\tau}, \tau \neq t\}| x_t)\cdot P(x_t)
\label{eq:productrule}
\end{equation}
And the sum rule to the denominator to define $P(X_{(t-1)})$ as the marginal distribution of $P(X_t)$:
\begin{equation}
    P(\{X_{\tau}, \tau \neq t\}) = \sum_T P(\{X_{\tau}, \tau \neq t\}, x_t) = \sum_T P(\{X_{\tau}, \tau \neq t\}| x_t = T)\cdot P(x_t = T)
    \label{eq:sumrule}
\end{equation}
where T comprehends all the possible states of $x_t$.\par
The Bayes Rule is the result of applying these two properties to the definition of conditional probability (equation \ref{eq:xtgivenxt-1}): 
\begin{equation}
    P(x_t|\{X_{\tau}, \tau \neq t\}) = \frac{P(X_t)}{P(\{X_{\tau}, \tau \neq t\})} = \frac{P(x_t) \cdot P(\{X_{\tau}, \tau \neq t\}|x_t)}{\sum_{T} P(x_t = T)\cdot P(\{X_{\tau}, \tau \neq t\}|x_t = T)}
    \label{eq:Bayes}
\end{equation}
% QUESTION OF THE DENOMINATOR
The general form of the Bayes Theorem says that \emph{the posterior probability of an event is proportional to the prior probability of that event times the likelihood of the observation conditioned to that event}. In other words, if the probability of a given variable (or set of variables) $P(\{X_{\tau}, \tau \neq t\})$ is fixed, the posterior probability (equation \ref{eq:Bayes}) can be expressed as a proportional factor of the numerator.\par
Using the previous example that tracks the value of a pixel over 3 consecutive frames, the value of the pixel at time frame 4 will be given by:

\begin{equation}
    P(x_4|X_3) \propto P(x_4) \cdot P(X_3|x_4) 
    \label{eq:Bayesgeneral}
\end{equation}

In a more general form, the conditional probability of a state $x_t$ given a set of previous observations from $x_1$ to $x_{(t-1)}$:

\begin{equation}
    P(x_t|\{X_{\tau}, \tau \neq t\}) \propto P(x_t) \cdot P(\{X_{\tau}, \tau \neq t\}|x_t) 
    \label{eq:Bayesextended}
\end{equation}

\noindent\fbox{%
    \parbox{\textwidth}{%
The Naive Bayes assumption states that the features (observations $X_{(t-1)}$) are conditionally independent given the class label ($x_t$)\citep{Murphy2012MachineL}. 
    }%
}
Applying the Naive Bayes assumption of independence allows to exploit the second term of equation \ref{eq:Bayesextended} into '$t-1$' different terms:

\begin{equation}
    P(\{X_{\tau}, \tau \neq t\}|x_t) = P(x_1|x_t) \cdot P(x_2|x_t), ..., P(x_{(t-1)}|x_t)\\
    P(x_t|\{X_{\tau}, \tau \neq t\}) \propto P(x_t)\prod_{n=1}^{(t-1)} P(x_n|x_t) 
    \label{eq:Bayesfinal}
\end{equation}

\begin{figure}[ht]
    \centering
    \includegraphics[scale=0.35]{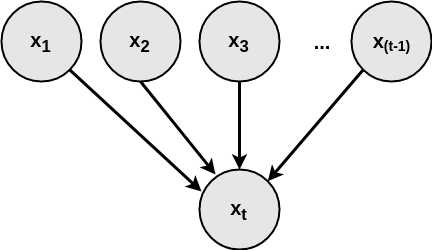}
    \caption{Graphical representation of the final form Naive Bayes classifier. It is based on the Naive Bayes assumption that states that: the observations ($x_1, ... x_{(t-1)}$) are conditionally independent given the value of $x_t$.}
    \label{fig:NaiveBayes}
\end{figure}

Equation \ref{eq:Bayesfinal} defines a model that predicts the value for the state $x_t$ for a set of observed states $X_{(t-1)}=(x_1,x_2,...,x_{(t-1)})$. It is the final form of the Naive Bayes classifier, which as a consequence of the Naive Bayes assumption do not capture dependencies between each of the observed states in $X_{(t-1)}$ (figure \ref{fig:NaiveBayes}). Even though this conditional independence assumption might sound unrealistic for real case scenarios, empirical results have shown a good performance in multiple domains with attribute dependencies \citep{Domingos1996BeyondIC}. These positive findings can be explained due to the loose relation between classification and probability estimation: 'correct classification can be achieved even when the probability estimates used contain large errors' \citep{Domingos1996BeyondIC}.\par

Markov Models and Hidden Markov Models do not make any assumptions about the in-dependency of the variables and will be illustrated next.

\subsubsection{Markov Chains}\label{sec:MM}
\begin{figure}[ht]
    \centering
    \includegraphics[scale=0.35]{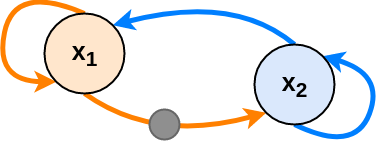}
    \caption{Graphical representation of a Markov Model with two possible states: $x_1$ and $x_2$. The connections between states represent the possible paths that the current state can follow. The values that condition each path are usually contained in a transition table (table \ref{tab:transition matrix}). The dark circle is current state transitioning from $x_1$ to $x_2$. Image adaptation from: \cite{MarkovChains}.}
    \label{fig:Markov Chain}
\end{figure}

Markov chains or Markov Models (MM) are stochastic generative models for dynamic systems that follow the Markov property. Markov's property states that the future state of a variable depends only on the current observation (there is only dependence between adjacent periods). In the framework of video processing, Markov's property can be interpreted as: the value of a pixel in the present is only dependent on its immediate past (figure \ref{fig:Markov Chain}).\par
 
Using probabilistic notation, Markov's property can be applied as:
\begin{equation}
    P(x_t|\{X_{\tau}, \tau \neq t\})=P(x_t|x_{(t-1)}) \label{eq:MM}
\end{equation}
Where $\{X_{\tau}, \tau \neq t\}$ contains all the previous states from $x_1$ to $x_{(t-1)}$. The resulting joint probability of a Markov chain like the one present in figure \ref{fig:Markov Chain}, is defined as:
\begin{equation}
    P(X_t)=P(x_1)P(x_2|x_1)P(x_3|x_2)...=p(x_1)\prod_{t=2}^T P(x_t|x_{(t-1)})
    \label{eq:1-oMM}
\end{equation}
In discrete-MM, the variables can only take certain values from a set of possible states that differ from each other. For a set of N possible states, there is a N by N transition matrix that contains the probabilities of transitioning between states. Figure \ref{fig:Markov Chain} shows an example of a MM with two possible states $x_1$ and $x_2$, the transition probabilities that define this MM can be found in table \ref{tab:transition matrix}. 
 
\begin{table}[ht]
    \centering
    \begin{tabular}{|c|c|c|}
        \hline
         & $x_1$ & $x_2$ \\
        \hline
        $x_1$ & $P(x_1|x_1)$ & $P(x_2|x_1)$\\
        \hline
        $x_2$ & $P(x_1|x_2)$ & $P(x_2|x_2)$\\
        \hline
    \end{tabular}
    \caption{Transition matrix of a Markov Model with two possible states ($x_1$ and $x_2$).}
    \label{tab:transition matrix}
\end{table}

In a discrete stochastic process, the rows of a transition probability matrix have to sum up to one, which means that a state has a finite amount of possible states. The values inside the transition matrix can be: given; calculated gathering samples from the process, doing a statistical analysis of the data and assuming that the process follows a certain distribution; or approximated using probability distribution approximation methods such as the Metropolis-Hastings algorithm, that assumes an initial probability distribution and through several iterations it moves it closer to the real distribution \citep{Murphy2012MachineL}.\par
\begin{figure}[ht]
    \centering
    \includegraphics[scale=0.35]{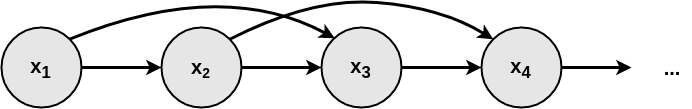}
    \caption{Graphical representation of a second order Markov chain. Image adaptation from: \cite{Murphy2012MachineL}}
    \label{fig:2-oMM}
\end{figure}

The strong assumption made in equation \ref{eq:MM} can be relaxed by adding dependence with more than one past states, transforming the MM into a k-order Markov chain. A second order Markov chain is illustrated in figure \ref{fig:2-oMM}.\par
The corresponding joint probability of a second-order Markov chain follows the next equation:
\begin{equation}
    P(X_t)=P(x_1,x_2)P(x_3|x_1,x_2)P(x_4|x_2,x_3)...=p(x_1,x_2)\prod_{t=3}^T P(x_t|x_{(t-1)}, x_{(t-2)})
    \label{eq:2-oMM}
\end{equation}
 
Equations \ref{eq:MM} and \ref{eq:2-oMM} can be applied where the for processes where the state of the system can be directly observed. However, in many applications, the state is not directly observable, these are called Hidden Markov Models and will be defined next.

\subsubsection{Hidden Markov Model}\label{sec:HMM}
Hidden Markov models (HMM) also belong to the stochastic generative models category. They differ from Markov chains because the state variables '$Z_t = {(z_1, z_2, ... z_t)}$' are not anymore directly accessible, i.e. hidden variables, and only the variables '$X_t = {(x_1, x_2, ... x_t)}$' are observable.
\begin{figure}[ht]
    \centering
    \includegraphics[scale=0.35]{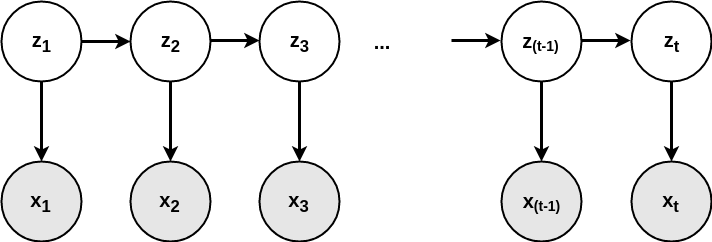}
    \caption{Graphical representation of a first order hidden Markov model. Image adaptation from: \cite{Murphy2012MachineL}}
    \label{fig:HMM}
\end{figure}
 
Figure \ref{fig:HMM} shows the representation of a first order HMM. There are two equations necessary to define this HMM. The relation between observable variables $X_t$ and hidden processes $Z_t$; and the relation between hidden processes with each other:

\begin{equation}
    P(x_{t}|\{X_{\tau}, \tau \neq t\},Z_t) = P(x_t|z_t)  \label{eq:HMM-xz}\\
    P(z_{t}|\{Z_{\tau}, \tau \neq t\}) = P(z_{t}|z_{(t-1)}) \label{eq:HM-zz}
\end{equation}

Resulting in the following joint distribution:  
\begin{equation}
    P(Z_{t},X_{t}|Z_{t}) = P(X_{t}|Z_{t})P(Z_{t}|\{Z_{\tau},t-1\leq \tau < t\})\\
    = P(z_1, x_1|z_1)\prod_{t=2}^T P(x_{t}|z_{t})P(z_{t}|z_{(t-1)}) \label{eq:HMM-jointzx}\\
\end{equation}
 
The probabilities that relate hidden states '$z_t$' (equation \ref{eq:HM-zz}) are called transition probabilities, while the probabilities that associate hidden processes with observable variables '$x_t$' (equation \ref{eq:HMM-xz}) are the emission probabilities. Both of them can be calculated in an analogous way to the transition probabilities for the Markov chains (table \ref{tab:transition matrix}).\par
Markov Models and Hidden Markov Models, although more general than the Naive Bayes Classifier, are also limited by definition. Each state is defined only to be affected by a finite number of previous states '$k$' and the effect of any other states happening before '$t-k$' is assumed to be encoded in this period, this limitation is often described as a short-term memory problem. Trying to find patterns in the sequence to determine the k-gram dependencies beforehand can help to alleviate this issue \citep{Conklin-MarkovLimitations}.\par

This section has introduced different methods that are used to model sequential information from the probabilistic theory point of view. Next section introduces different approaches used to overcome temporal context using deep learning architectures.

\subsection{Video Modeling: Deep Learning Architectures}
Considering videos as sequences of static images, this section can serve as an introduction to different approaches used to add temporal context to the analysis of videos.

\subsubsection{Gated Recurrent Flow Propagation - GRFP}\label{sec:grfp}
Seeking to solve the semantic segmentation inconsistency characteristic of evaluating video segmentation with individual image segmentation methods, \cite{Nilsson2016SemanticVS} announced a method that combines nearby frames and gated operations for the estimation of a more precise present time segmentation.\par

\begin{figure}[ht]
    \centering
    \includegraphics[scale=0.55]{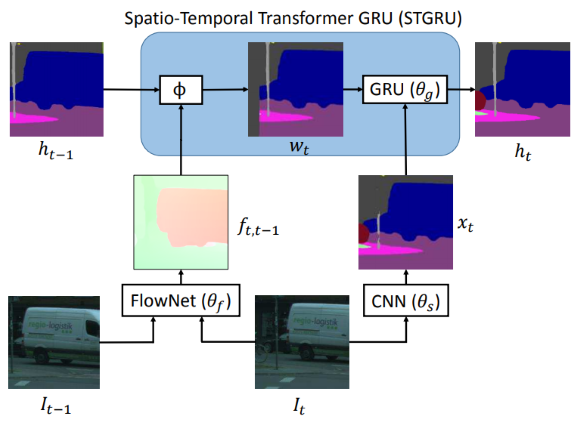}
    \caption{Overview of the Spatio-Temporal Transformer Gated Recurrent Unit. Pairs of raw input images are used to calculate the optical flow of the image (FlowNet). This optical flow is then combined with the semantic segmentation of the previous frame, obtaining a prediction of the present segmentation (blue box). A segmentation map of the present frame is then passed together with the prediction to a GRU unit that combines them based on the sequence. Image source: \cite{Nilsson2016SemanticVS}}
    \label{fig:GRFP}
\end{figure}

The Spatio-Temporal Transformer GRU (STGRU) in \cite{Nilsson2016SemanticVS}, is a network architecture that adopts multi-purpose learning methods with the final purpose of video segmentation. Figure (\ref{fig:GRFP}) shows a scheme of the STGRU architecture.
Inside of the STGRU, FlowNet is in charge of calculating the optical flow for N consecutive frames. A wrapping function ($\phi$) uses this optical flow to create a prediction of the posterior frame. Later, a GRU compares the discrepancies between the estimated frame ($w_t$) and the current frame evaluated by a baseline semantic segmentation model ($x_t$), keeping the areas with higher confidence while reseting the rest of the image.\par

\begin{figure}
    \centering
    \includegraphics[scale=0.75]{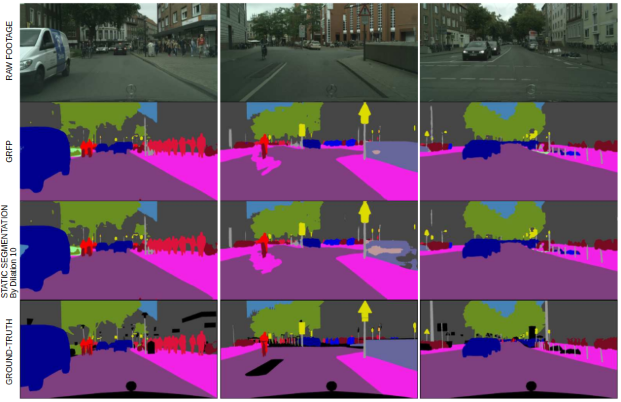}
    \caption{Image adaptation from Nelsson et al. \citep{Nilsson2016SemanticVS}. In this image it can be seen a comparison between: GRFP method, Static Semantic Segmentation and the groundtruth segmentation. From left to right, GRFP achieves a segmentation improvement for the left-car, the right-wall and the left-pole.}
    \label{fig:GRFP-results}
\end{figure}
The STGRU presented in \cite{Nilsson2016SemanticVS} was evaluated both quantitatively and qualitatively (fig. \ref{fig:GRFP-results}), exhibiting a high performance compared with other segmentation methods.

\subsubsection{Other architectures for video applications}
The current trend for semantic video segmentation models consists on combining multipurpose neural networks (sequence modelling with feature extraction networks) into advanced models capable of efficiently performing this task.\par
Some other video segmentation architectures include: 
\begin{itemize}
    \item Feature Space Optimization for Semantic Video Segmentation \citep{Kundu2016FeatureSO}.
    \item Multiclass semantic video segmentation with object-level active inference \citep{Liu2015MulticlassSV}.
    \item Efficient temporal consistency for streaming video scene analysis \citep{Miksik2013EfficientTC}.
    % \item Semantic Video Segmentation: Exploring Inference Efficiency uses CRF MAYBE NOT INCLUDE THIS ONE AS IT ONLY EXPLORES DE EFFICIENCY
\end{itemize}
Semantic image segmentation is not the only application in computer vision that can benefit from leveraging temporal context, tracking also use temporal analysis tools to achieve a better performance. The main reasons why temporal context is necessary for tracking are to guarantee the detection of the object even through occlusion and to reduce the number of identity switches (during multiple object tracking). These kind of applications are very common on surveillance or sport events. In 2017, \citep{Wojke2017SimpleOA} combined image appearance information with other tracking methods (Simple Online Real-time Tracker \citep{Bewley2016SimpleOA}) based on Kalman Filter and the Hungarian algorithm to obtain state-of-the-art detection at high rates (40Hz).\par

\section{Analysis}
This section provides a detailed description of the problem and the materials that will be used for the study. 

\subsection{Domain Analysis}

\subsubsection{Semantic segmentation for Autonomous cars}
\begin{figure}[ht]
    \centering
    \includegraphics[scale=0.4]{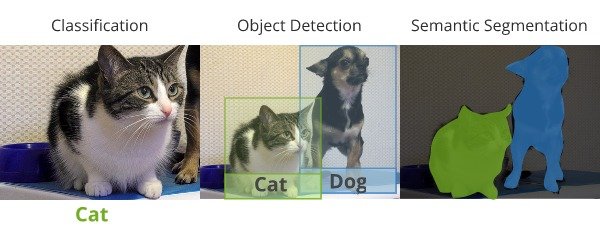}
    \caption{Figure comparison of the three computer vision classification applications. From left to right: image classification (global classification), object detection (local classification) and semantic segmentation (pixel-level classification). Image source: \cite{catdogimage}}
    \label{fig:catdogimage}
\end{figure}

Semantic segmentation is considered one of the hardest computer vision applications. It differs from image classification or object detection in how the classification is performed (figure \ref{fig:catdogimage}). Image classification models classify the image globally, they assign a label to the whole image (e.g. cat or dog image classifier). Object detection models look for patterns in the image and assign a bounding box to the region of the image that is more likely to match with the target description (it provides classification and location within the image). And semantic segmentation produces pixel-level classification of an image; it describes each pixel of the image semantically, providing a more insightful description of how the image is composed than the other two methods.\par
\begin{figure}[ht]
    \centering
    \includegraphics[scale=0.4]{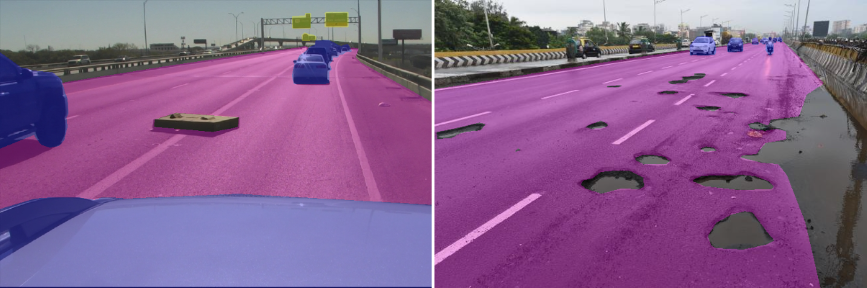}
    \caption{Adverse situations that can be solved using semantic image segmentation. Object detection models can only to detect objects that the model is familiar with. However, it is very difficult to create a dataset that includes all the possible types of obstacles, imperfections or debris that may appear in the road. Semantic image segmentation aims to achieve perfect definition of the image even when the objects are unknown.}
    \label{fig:debris}
\end{figure}

Humans are very good at segmentation of images, even without knowing what the objects are. This is the main reason why semantic image segmentation is necessary for autonomous navigation. Although other detection models are able to classify obstacles and locate them in the space, they can only find the obstacles they have previously seen. E.g. an obstacle detector used to avoid pedestrians in autonomous cars, it will only be able to alert the vehicle in the presence of pedestrians (it was just trained to learn how the pedestrian category is modeled). However, the type of obstacles that can be found in a undefinable driving scenario (it covers any object in any kind of shape and it is not feasible to create a data set that covers for all), this is the reason why semantic image segmentation is present in autonomous navigation. An ideal semantic image segmentation model will be able to define the boundaries of any objects, even when these objects have not been previously 'seen' (figure \ref{fig:debris}). Apart from being a good obstacle detector, a perfect semantic image segmentation model has the ability to store these previously unseen objects, tag them and use them to re-train the network and improve the accuracy.

There are some requirements that need to be present when applying semantic segmentation into autonomous navigation. The vehicle receives the data as a stream of images (it does not count with a video of the route beforehand) and it has to perform inference in real-time. The model should be very sensible on the detection of obstacles, e.g in an ambiguous situation where the segmentation of the road is not perfectly clear, due to imperfections or the presence of objects, the classification of obstacles must prevail.

\subsubsection{Software analysis}\label{sec:software}
The most common programming languages used for computer vision and deep learning applications are Python and C++. The former is preferred on the research scope, while the latter is mainly used in commercial applications. Apart from the programming languages, there are different frameworks that provide the developer with the tools required to handle big amounts of data: Theano, PyTorch, TensorFlow or Keras are some of the frameworks compatible with both Python or C++.\par
Although the programming language and deep learning framework affect the performance of the application, the final performance only depends on the implementation algorithm (semantic image segmentation model). As a matter of preference, this study is developed using Python and Tensorflow.\par

\subsubsection{Semantic image segmentation model selection}\label{sec:modelSelection}
\begin{figure}[ht]
    \centering
    \includegraphics[scale=0.55]{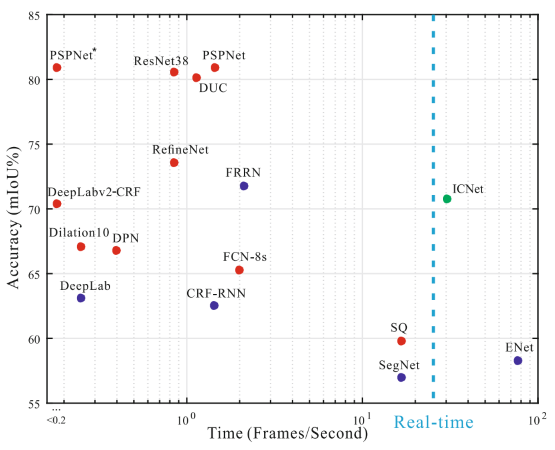}
    \caption{Classification of different semantic image segmentation models according to inference speed and accuracy (mIOU) on Cityscapes test set \citep{Cordts2016Cityscapes}. It can be observed how faster models (bottom-right corner) usually achieve a lower level of accuracy than slower ones (upper-left corner). Image source: \cite{Zhao2017ICNetFR}.}
    \label{fig:SSclassification}
\end{figure}

Depending on its inner structure, the different semantic image segmentation models are able to obtain different levels of segmentation accuracy and inference speed. Figure \ref{fig:SSclassification}, although it is not up to date, shows some of the available possibilities arranged by accuracy and inference speed on Cityscapes dataset \citep{Cordts2016Cityscapes}. Cityscapes is a large-scale urban-scene dataset that contains high resolution fully annotated segmentation images for semantic segmentation applications (section \ref{sec:cityscapes}). \par
In figure \ref{fig:SSclassification}, the inference speed was measured by counting the amount of frames that the segmentation model is able to process each second. The mIoU (mean Intersection over Union) measures the mean accuracy of all the frames processed at each speed. As a result, the upper-left corner contains the most accurate (but slower, $\sim0-1$ frames per second) models while the less accurate (but faster, $\sim10-100$ frames per second) are grouped on the bottom-right corner.\par
Later in the same year of the release of the study in figure \ref{fig:SSclassification}, Chen et al. \citep{Chen2017RethinkingAC} in their paper \emph{Rethinking Atrous Convolution for Semantic Image Segmentation}, introduced DeepLabv3, a new iteration of the DeepLab model series that became the state-of-the-art for semantic image segmentation models on Cityscapes test set (table \ref{tab:modelSelection}).\par
In order to continue with state-of-the-art efficiency, DeepLabv3 \citep{Chen2017RethinkingAC} with pretrained-weights on Cityscapes dataset \citep{Cordts2016Cityscapes} is choosen as the baseline model for this study.
\begin{table}[ht]
    \centering
    \begin{tabular}{c|c}
    \hline
    \textbf{Method} & \textbf{mIOU}\\
    \hline
    DeepLabv2-CRF \citep{Chen2016DeepLabSI} & 70.4\\
    Deep Layer Cascade \citep{Li2017NotAP} & 71.1\\
    ML-CRNN \citep{Fan2016MultiLevelCR}& 71.2\\
    Adelaide\_context \citep{Lin2015EfficientPT}& 71.6\\
    FRRN \citep{Pohlen2016FullResolutionRN}& 71.8\\
    LRR-4x \citep{Ghiasi2016LaplacianPR}& 71.8\\
    RefineNet \citep{Lin2016RefineNetM}& 73.6\\
    FoveaNet \citep{Li2017FoveaNetPU}& 74.1\\
    Ladder DenseNet \citep{Krapac2017LadderStyleDF}& 74.3\\
    PEARL \citep{Jin2016VideoSP}& 75.4\\
    Global-Local-Refinement \citep{Zhang2017GlobalresidualAL}& 77.3\\
    SAC\_multiple \citep{Zhang2017ScaleAdaptiveCF}& 78.1\\
    SegModel \citep{Shen2017SemanticSV}& 79.2\\
    TuSimple\_Coarse \citep{Wang2017UnderstandingCF}& 80.1\\
    Netwarp \citep{Gadde2017SemanticVC}& 80.5\\
    ResNet-38 \citep{Wu2016WiderOD}& 80.6\\
    PSPNet \citep{Zhao2016PyramidSP}& 81.2\\
    \hline
    \textbf{DeepLabv3} \citep{Chen2017RethinkingAC}& \textbf{81.3}
    \end{tabular}
    \caption{Table comparison of the performance of different semantic image segmentation models on the Cityscapes dataset \citep{Cordts2016Cityscapes}. Table adaptation: \cite{Chen2017RethinkingAC}.}
    \label{tab:modelSelection}
\end{table}

\subsubsection{DeepLabv3}\label{sec:deeplabv3}
DeepLabv3 \citep{Chen2017RethinkingAC}, developed by Google, is the latest iteration of the DeepLab model series for semantic image segmentation-- previous versions: DeepLabv1 \citep{Chen2014SemanticIS} and DeepLabv2 \citep{Chen2016DeepLabSI}.\par
DeepLab is based on a fully convolutional layer architecture (FCN) that employs atrous convolution with upsampled filters to extract dense feature maps and capture long range context \citep{Chen2017RethinkingAC}. \citep{Shelhamer2014FullyCN} showed how powerful convolutional networks are at elaborating feature models and defined a FCN architecture that achieved state-of-the-art performance for semantic segmentation on the PASCAL VOC benchmark \citep{Everingham2009ThePV}. Another of the advantages of using FCN is that the architecture is independent of the input size, they can take input of arbitrary size and produce correspondingly-sized output \citep{Shelhamer2014FullyCN}. In contrast, architectures that combine Convolutional Networks (for feature extraction) with Fully-Connected Conditional Random Fields (for classification)\citep{Chen2014SemanticIS, Chen2016DeepLabSI} are designed for a fixed input size, as a result of the particular size necessary to pass through these classification layers.\par
\begin{figure}[ht]
    \centering
    \includegraphics[scale=0.55]{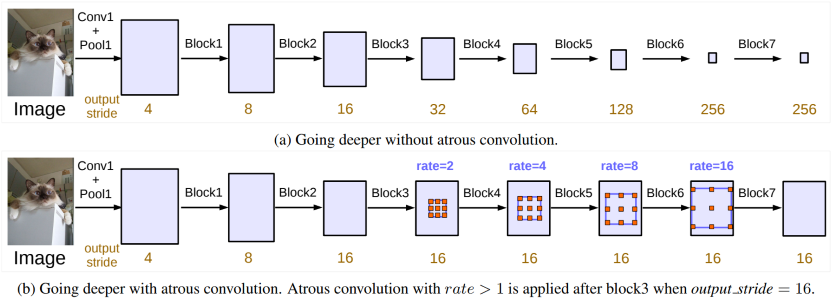}
    \caption{This figure compares the effect of consecutive pooling or striding to the feature map. (a) Shows an example where the feature map in the last layers is condensed to a size 256 times smaller than the input image, this is harmful for semantic segmentation since detail information is decimated \citep{Chen2017RethinkingAC}. (b) Applies atrous convolution to preserve the output stride obtaining equivalent levels of abstraction \citep{Chen2017RethinkingAC}. Image source: \citep{Chen2017RethinkingAC}}
    \label{fig:outputStride}
\end{figure}

One main limitation of solving semantic segmentation using Deep Convolutional Neural Networks (DCNNs) are the consecutive pooling operations or convolution striding that are often applied into the DCNNs, consequently reducing the size of the feature map. These operations are necessary in order to increasingly learn new feature abstractions \citep{Scherer2010EvaluationOP}, but may impede dense prediction tasks, where detailed spatial information is desired. Chen et al. \citep{Chen2017RethinkingAC} suggest the use of 'atrous convolution' as a substitute of the operations that reduce the size of the input (figure \ref{fig:outputStride}).\par
\begin{figure}[ht]
    \centering
    \includegraphics[scale=0.45]{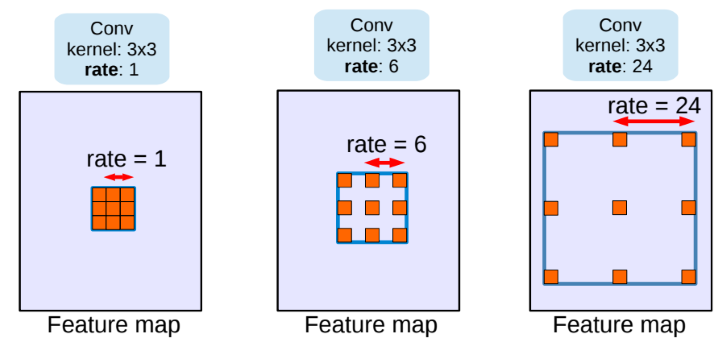}
    \caption{Atrous convolution of a filter with kernel size 3x3 at different stride rates. The dilation rate determines the space in between the different cells of the kernel. A rate=1 corresponds to the standard convolution, a rate=6 means that weights of the kernel are applied every sixth element (gap of 5 units), and so on. Image source: \citep{Chen2017RethinkingAC}.}
    \label{fig:atrousConv}
\end{figure}

Atrous convolution, is also known as a dilated convolution. Apart from the kernel size, dilated convolutions are specified by the dilation rate, that establishes the gap between each of the kernel weights. A dilation rate equal to one corresponds to a standard convolution, while a dilation rate equal to two means that the filter takes every second element (leaving a gap of size 1), and so on (figure \ref{fig:atrousConv}). The gaps between the values of the filter weights are filled by zeros, the term 'trous' means holes in French. \citep{Chen2014SemanticIS, Yu2015MultiScaleCA, Chen2016DeepLabSI} show how effective the application of dilated convolution is in maintaining the context of the features.\par
\begin{figure}[ht]
    \centering
    \includegraphics[scale=0.4]{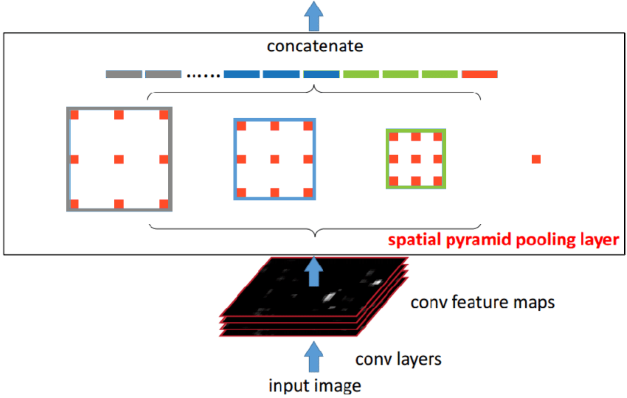}
    \caption{Graphical representation of ASPP. Atrous convolution with different dilation rates is applied on the same feature map, the result of each convolution is then concatenated and passed to the next layer. Spatial pyramid pooling is able to capture feature context at different ranges\citep{Grauman2005ThePM, Lazebnik2006BeyondBO, Chen2016DeepLabSI}. Image adaptation: \citep{asppfigure}}
    \label{fig:aspp}
\end{figure}

\begin{figure}[ht]
    \centering
    \includegraphics[scale=0.52]{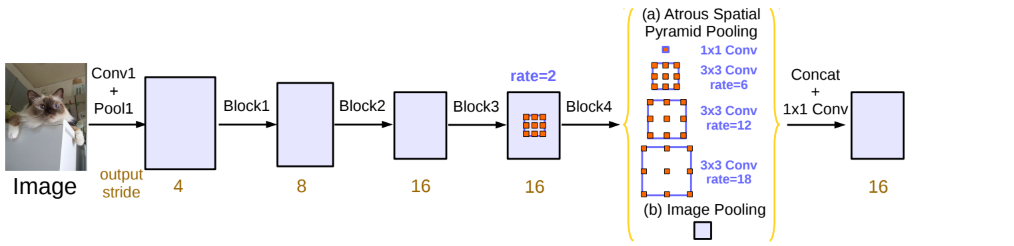}
    \caption{DeepLabv3 architecture. The first 3 blocks are a replica of the last block of the original residual neural network \citep{He2015DeepRL}. The following blocks incorporate the use of atrous convolution and ASPP, which stops the output stride reduction to 16, diminishing the negative effects of consecutive pooling. Image source: \citep{Chen2017RethinkingAC}}
    \label{fig:deeplabv3}
\end{figure}

A second limitation faced by semantic image segmentation models is that they have to detect objects at multiple scales. This is a problem when using regular sized filters (normally 3x3) because they can only 'see' in regions of 9 pixels at a time, which makes it very difficult to capture the overall context of big objects. DeepLabv3 \citep{Chen2017RethinkingAC} employs Atrous Spatial Pyramid Pooling (ASPP) to overcome this issue, it consists on applying atrous convolution with different dilation rates over the same feature map and concatenate the results before passing it into the next layer figure \ref{fig:aspp}. This approach helps capturing feature context at different ranges without the necessity of adding more parameters into the architecture (larger filters) \citep{Grauman2005ThePM, Lazebnik2006BeyondBO, Chen2016DeepLabSI}.
Figure \ref{fig:deeplabv3} shows the final architecture of DeepLabv3. Blocks 1 to 3 contain a copy of the original last block in ResNet \citep{Chen2017RethinkingAC}; which consists of six layers with 256 3x3 kernel convolution filters (stride=2), batch normalization right after each convolution and skip connections every 2 layers \citep{He2015DeepRL}. Block 4 is equivalent to the first 3 but it applies atrous convolution with dilation rate of 2 as a substitute of downsampling the image with convolutions of stride 2, maintaining the output stride to 16. The next block applies ASPP at different rates and global average pooling of the last feature map, all the results of this block are then concatenated and passed forward. The resulting features from all the branches are then concatenated and passed through a 1x1 convolution before the final 1x1 convolution that generates the final logits \citep{Chen2017RethinkingAC}.
The output image consists on a HxWxC matrix where H and W correspond to the height and width of the output image and C is the number of categories in the dataset. Every pixel is assigned a real number for each category, that represents the likelihood (or logits) of that pixel belonging to each category, this is called the score map. The score map is then reduced by means of an argmax operation that determines the index of the category with the highest likelihood, obtaining the semantic segmentation map.

\subsubsection{Cityscapes dataset}\label{sec:cityscapes}
Cityscapes is the state-of-the-art dataset for urban scene understanding. It was created out of the lack of available datasets that adequately captured the complexity of real-world urban scenes. Despite the existence of generic datasets for visual scene understanding such as PASCAL VOC \citep{Everingham2009ThePV}, the authors of Cityscapes claim that "\textit{serious progress in urban scene understanding may not be achievable through such generic datasets}" \citep{Cordts2016Cityscapes}, referring to the difficulty of creating a dataset that can cover any type of applications.\par
\begin{figure}[ht]
    \centering
    \includegraphics[scale=0.4]{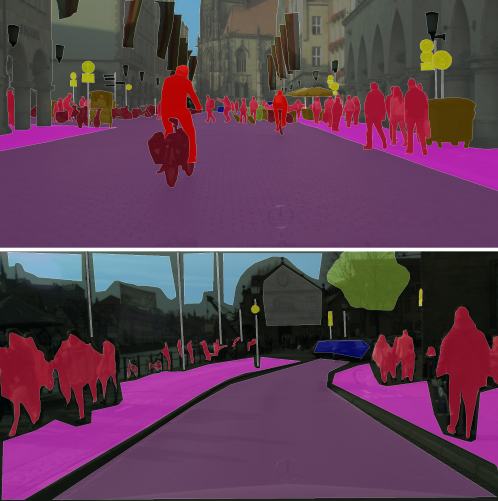}
    \caption{Different types of annotations present in the Cityscapes dataset \citep{Cordts2016Cityscapes}. Upper figure shows an example of a densely annotated image (richer in detail). Bottom figure shows an example of a coarsely annotated image (lower level of detail). Image source: \citep{cityscapesweb}.}
    \label{fig:fineVScoarse}
\end{figure}

Nonetheless, Cityscapes is not the only dataset of its kind. Other datasets such as CamVid \citep{BrostowSFC:ECCV08}, DUS \citep{DUS} or KITTI \citep{Geiger2013IJRR} also gather semantic pixel-wise annotations for the application in autonomous driving. Cityscapes is the largest and most diverse dataset of street scenes to date \citep{Cordts2016Cityscapes}, it counts with 25000 images (figure \ref{fig:fineVScoarse}) from which 5000 are densely annotated (pixel-level annotation) while the remaining 20000 are coarsely annotated (using bounding polygons, which offers a lower level of detail). Compared to the other datasets purposed for autonomous driving, Cityscapes has the largest range of traffic participants (up to 90 different labels may appear in the same frame)\citep{Cordts2016Cityscapes} and has the largest range for object distances, covering objects up to 249 meters away from the camera \citep{Cordts2016Cityscapes}.\par

All these characteristics make the Cityscapes dataset the most challenging urban-scene understanding benchmark to date, "\textit{algorithms need to take a larger range of scales and object sizes into account to score well in our benchmark}"\citep{Cordts2016Cityscapes}. Yet there are some limitations that need to be considered when evaluating the performance of a model that has been trained using this dataset. Cityscapes only captures urban areas (inner-city) of cities primarily from Germany or neighbouring countries \citep{Cordts2016Cityscapes}, which may turn in a low efficiency when applied to highways, rural areas or other countries (due to differences in the architectures). The original images were taken during spring and summer seasons and do not cover situations with adverse weather conditions and or poor illumination. More information about the composition and a statistical analysis of this dataset can be found in \citep{Cordts2016Cityscapes}.

% \subsection{DeepLabv3 on Cityscapes performance analysis}\label{sec:baseline}
% Failure modes: completely missed classifications, partial segmentation, false classifications
% Flickery effect,
% Changes in illumination
% ADD IMAGES
% From now on an instance of DeepLabv3 pre-trained on the Cityscapes dataset will be regarded as the baseline model.\par

\subsection{Methodology}\label{sec:methodology}
After several test runs of the  baseline model on image sequences, it was observed that the production of wrong classifications (completely or partially missed object's classification) is a transitory effect. When applied on a sequence of images, the baseline model is usually able to detect most of the objects producing segmentation of different qualities on each frame. Differences in the lighting conditions or noise in the image may be the cause of this variation from frame to frame. However, \textbf{this is an effect that can be exploited in order to achieve a better segmentation}.\par
The next conclusion came after analyzing a moving object over consecutive frames. The moving object was recorded using a regular camera (at 30fps), it was noticed how the displacement of the subject from frame to frame was very small, depending on its relative speed with respect to the motion of the camera and its distance from the camera. This small displacement produced by objects moving at relatively low-medium speeds (walking person, moving bike or moving car) can be used as a motivation that the segmentation of neighbouring frames can be combined in order to obtain a more accurate segmentation of the present. In the next sections, this concept is regarded as the Image Buffer approach, figure \ref{fig:personSegmentationOverlapAnalysis} shows an illustration of and Image Buffer of size 2: it holds 2 frames from the past and merges them to the segmentation of the present frame.\par
Apart from a straightforward combination of the pixel-classification output of neighbouring time frames (Image Buffer, section \ref{sec:imageBuffer}), a second approach that computes a weighted combination of the pixel-classification logits (section \ref{sec:deeplabv3}) produced by the baseline model will be introduced next.\par 
As explained in section \ref{sec:deeplabv3}, the way in which DeepLab assigns the final classification labels to each pixel is by a previous calculation of a C-dimensional score array for each pixel, that together form a HxWxC scoremap for the input image (C is the number of possible categories, $19$ in the case of Cityscapes; H and W correspond to the height and width of the input image respectively). The C-dimensional array contains the likelihood of each pixel to belong to each one of the possible categories of the data set. The final pixel-labels are assigned by reducing the C-dimensional array of each pixel into one value that represents the index of the maximum value of the array, this is done by means of an argmax operation.\par
The weighted combination of the classification scores as an approximated version of a conditional probability (section \ref{sec:conditionalprob}) is the second approach that will be tested in the following sections. This method is referred as Attention Module (section \ref{sec:attentionModule}).\par

\subsubsection{Testing and evaluation}\label{sec:testingandev}
In order to stay truthful to the nominal conditions of baseline model, it would be necessary to test it using images with the same characteristics as the Cityscapes data set \citep{Cordts2016Cityscapes}, this is 2048x1024 pixel images. However, it was not possible to find video sources with the same resolution as the Cityscapes data set and a 1920x1080 pixel resolution was adopted for the different tests. Although the difference on the number of total pixels between both formats is less than 2 percent, using lower resolution images than the ones used for training the weights of the baseline model might have an effect on the final segmentation. The study of this issue has not been covered and is added as one of the limitations in the discussions section (section \ref{sec:limitations}).\par
The performance of both of the suggested approaches listed before, Image Buffer (\ref{sec:imageBuffer}) and Attention Module (\ref{sec:attentionModule}) as well as the baseline performance (\ref{sec:modelSelection}) will be evaluated both quantitatively and qualitatively.\par
It is necessary to count with a groundtruth label annotation data set to quantitatively measure the segmentation performance. However, groundtruth annotations for muli-label semantic video segmentation are very costly and no available data sets that covers this need were found. The Densely Annotated VIdeo Segmentation (DAVIS) 2016 benchmark \citep{Perazzi2016} was chosen as an approximation for this requirement. It is formed by 50 densely annotated single-object short sequences, from which only 10 are suitable for the evaluation of this exercise (due to compatibility with the Cityscapes preset categories). The DAVIS categories that will be used for this study are:\par
\begin{itemize}
    \item Breakdance-flare. A single person moving rapidly in the middle of the screen.
    \item Bus. A bus centered in the picture frame in a dynamic environment.
    \item Car-shadow. A car moving out from a shadow area.
    \item Car-turn. A car moving towards and outwards from the camera.
    \item Hike. A person moving slowly in the center of the frame.
    \item Lucia. A person moving slowly in the center of the frame.
    % \item Parkour. A person moving fast all over the frame.
    \item Rollerblade. A person moving fast from left to right of the frame.
    \item Swing. A person swinging back and forth and being temporarily occluded in the middle of the screen.
    \item Tennis. A person moving fast from right to left in the frame.
\end{itemize}
Since goal of this study is to improve the detection and the segmentation over time by reducing the number of missed classification and maintaining a consistent segmentation, the metrics used for the evaluation cover the temporal consistency and the accuracy.\par
A semantic image segmentation model that is consistent over time will produce a segmentation area with a smooth transition from frame to frame (depending on whether the tracked subject is moving or not). The segmentation area is calculated by counting the number of pixels classified with the target label at each time step. Afterwards, the frame-to-frame area variation is calculated as the difference of the area between consecutive frame pairs. A final computation of the standard deviation of these differences gives a global metric for the segmentation fluctuations (it is expected to obtain a lower number for more temporal consistent methods) that will be used for comparison of the different approaches.\par 
The accuracy is calculated using the Intersection Over Union (described in section \ref{sec:modelSelection}). And, in the same way as with the area, the frame-to-frame fluctuations of the accuracy are calculated as a comparison metric for all the approaches.\par
The qualitative evaluation consists on the observation and interpretation of the segmentation result of each one of the methods using different video sources.\par
% It is not possible to quantify and compare this performance due to the lack of groundtruth annotations for each one of the test videos.\par
The videos used for the qualitative evaluation are:
\begin{itemize}
\item Citadel - University of Twente
% \item Carr\'e - University of Twente
\item Carr\'e (modified) - University of Twente. One every ten frames was removed from the original clip to simulate a temporal occlusion or the faulty behavior of the camera sensor.
% \item Driving in Amsterdam. Video source: \cite{drivingDowntownAmsterdam}.
\item Driving in a tunnel. Video source: \cite{drivingDowntownAmsterdam}.
\item Driving under the rain. Video source: \cite{drivingUnderRain}.
\item Driving in the night. Video source: \cite{drivingNight}.
\item Driving in low sun. Video source: \cite{drivingLowSun}
\end{itemize}

\subsubsection{Remarks}
Videos are a very powerful source of information. In contrast with the analysis of pictures, videos provide objects with a temporal context as a series of frames that can be exploited to benefit segmentation. In order to do so, two different approaches will be evaluated: Image Buffer and Attention Module (section \ref{sec:methodology}).\par
These two approaches will build up on top of DeepLabv3 \citep{Chen2017RethinkingAC} pre-trained on the Cityscapes data set \citep{Cordts2016Cityscapes}, which is chosen as the baseline for this study due to its performance as the state-of-the-art semantic image segmentation model (section \ref{sec:modelSelection}).\par
The results will be evaluated both qualitatively and quantitatively in a series of videos chosen to cover a wide variety of scenarios (section \ref{sec:testingandev}). The metrics used for the comparison of the different approaches are chosen to cover both the accuracy and the temporal consistency of the predictions (section \ref{sec:testingandev}).\par
The machine learning framework and programming language are fixed together: Python and Tensorflow. Python allows for rapid script prototyping and debugging and counts with lots of libraries that make working with images and arrays very natural. Tensorflow includes large amounts of documentation online, a very vivid community and it is being constantly updated with new utilities that offer new possibilities for the use of Deep Learning (section \ref{sec:software}).\par

\section{Design and implementation}

\label{ch:four}
This section covers in detail both of the suggested approaches used to provide temporal context to the semantic image segmentation: Image Buffer (section \ref{sec:imageBuffer}) and Attention Module (section \ref{sec:attentionModule}).

\subsection{Approach I: Image Buffer}\label{sec:imageBuffer}
Image Buffer is the first of the approaches studied to solve for the semantic image segmentation models' temporal inconsistency. It is the result of analyzing an object's location from frame-to-frame and the quality of the segmentation provided by DeepLabv3 (section \ref{sec:modelSelection}) over consecutive frames.\par
\begin{figure}[ht]
    \centering
    \includegraphics[scale=0.5]{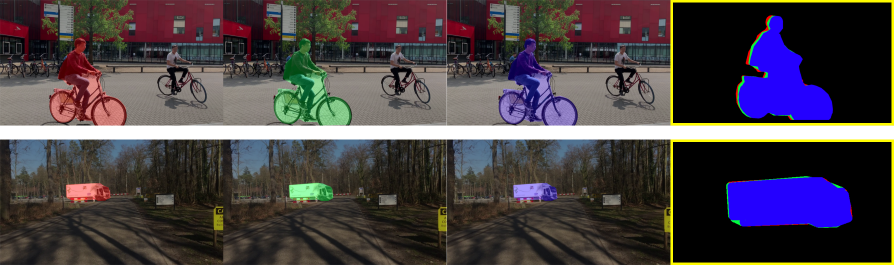}
    \caption{This figure compares the translation of a moving object during 3 consecutive frames. Each of the two series shown in this figure (bike and truck) correspond to 3 consecutive frames extracted from two videos recorded at 30 fps. The upper series (from left to right) show a subject moving in a close plane horizontal to the camera. The lower series (from left to right) show a truck moving in a far plane horizontal to the camera. The further to the right image for both series is a mask overlap of each target over the 3 consecutive frames (zoomed in), it represents the relative displacement over time of these objects for a video recorded at 30 fps. It is observed how the relative displacement in minimal for both examples. The bike series, introduces how the combination of the history of an object with larger relative displacements may introduce a 'ghosting effect' to the resulting image. Apart from being dependent on the recording speed (fps) and distance to the camera, the relative displacement is also dependent on the relative speed between camera and object (although a speed figure comparison is not elaborated, this was observed along the different tests for this thesis study).}
    \label{fig:displacementAnalysis}
\end{figure}

The first premise follows from figure \ref{fig:displacementAnalysis}. \textit{The translation of an object recorded at 30 fps over small batches of consecutive frames is negligible}.\par
\begin{figure}[ht]
    \centering
    \includegraphics[scale=0.8]{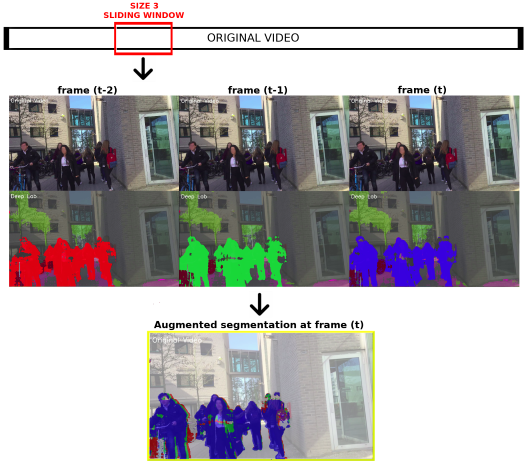}
    \caption{Conceptual figure that shows how the combination of the segmentation of a particular target (person in this case) of neighbouring frames can be beneficial for the segmentation at the present time. From top to bottom: 3 consecutive frames are extracted from the original video using a sliding window that covers the present frame and 2 other consecutive frames in the past (the extraction can be done for more than 3 frames); the baseline segmentation model (DeepLabv3 on Cityscapes) is applied to each one of these frames, the person label is highlighted in a different color (red, green, blue) for each frame in order to improve the visualization; the hypothetical augmented segmentation output is constructed by overlapping the segmentation of the 3 consecutive frames in the sliding window (bottom figure). It can be observed how the Augmented segmentation at frame (t) contains parts of the segmentation of each of the examined frames. The results show how the wholes in the segmentation of the man in the left and the woman in the middle at time (t) are 'filled' by the segmentation of frames at times (t-2) and (t-1).}
    \label{fig:personSegmentationOverlapAnalysis}
\end{figure}

The second premise follows from figure \ref{fig:personSegmentationOverlapAnalysis}. \textit{The combination of the segmentation in neighbouring frames helps completing the segmentation and avoiding temporal miss classifications}. In figure \ref{fig:personSegmentationOverlapAnalysis}, 3 consecutive segmented frames were overlap in order to obtain an 'augmented segmentation'. The result of this experiment showed how the figure of a person was completed using the information of the 3 involved frames.\par
\begin{figure}[ht]
    \centering
    \includegraphics[scale=0.55]{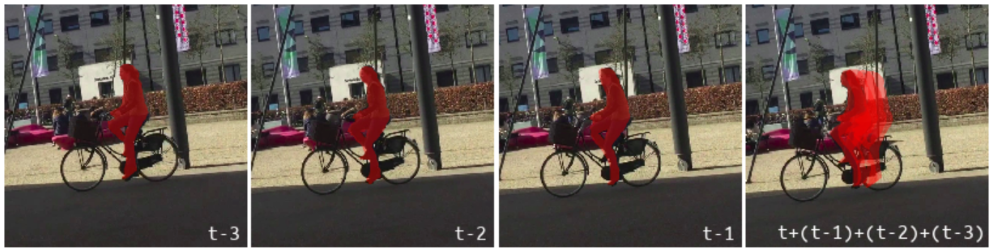}
    \caption{Block diagram of the Image Buffer approach (section \ref{sec:imageBuffer}). This block diagram depicts how the segmentation maps of 4 consecutive frames are combined into one (augmented segmentation).}
    \label{fig:imageBufferBlockDiagram}
    \includegraphics[scale=0.5]{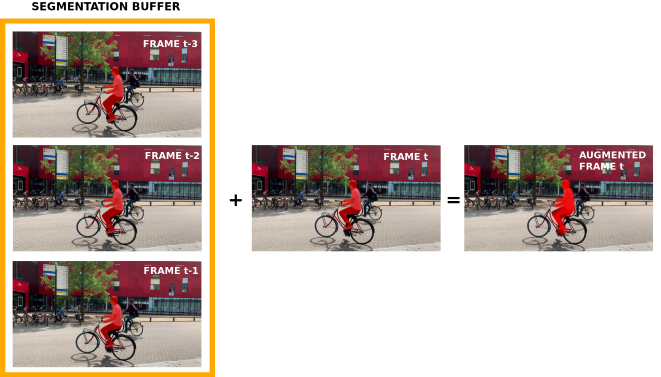}
    \caption{Conceptual representation of the Image-buffer approach. In this figure, the output is calculated as a combination of the segmentation in 3 consecutive past frames plus the segmentation of the present frame.}
    \label{fig:imageBufferRepresentation}
\end{figure}

The origins of this fault in segmentation could be of different nature, from temporal physical occlusion such as flying objects covering the camera lens, to sensor saturation due to sun glare, or information loss due to system performance problems. 
The approach works as follows: 
\begin{enumerate}
    \item The present time frame is extracted from the video feed and passed through DeepLabv3 that computes the segmentation map.
    \item The segmentation map is concatenated along with the segmentation of the 3 previous time frames contained in the Segmentation Buffer (figure \ref{fig:imageBufferBlockDiagram}).
    \item The segmentation of the 4 evaluated frames is combined obtaining the augmented segmentation for the present time (figure \ref{fig:imageBufferRepresentation}).
    \item Lastly, the Segmentation Buffer is updated by dropping the oldest frame and storing the last frame drawn from the original video.
\end{enumerate}
To delimit this approach it is necessary to define the size of the segmentation buffer, this follows from the observation made in figure \ref{fig:displacementAnalysis} that shows how the relative displacement on three consecutive frames is very small. For simplicity, the size of the image buffer is fixed to four: covers the present frame and three frames from the past.\par
This technique is only applied to a predefined subset of categories, the target categories. The reason for this discrimination is the concern towards the segmentation of the obstacles that can be found in the road. As a proof of concept, the target categories for the experiments are: 'person', 'rider', 'car' and 'bicycle'.\par
It is expected that this approach alleviates the defect of partially or completely missed classifications over the target categories. Nevertheless, the effect of this approach will only be apparent if the defect labels are missing for less than the size of the Image Buffer (if the labels are missing for longer than that, this approach is not able to recover them).\par
Overlapping past information for the inference of the present segmentation imposes a temporal lower bound to the object detection, at the same time, it reduces the time inconsistency alleviating the 'flickery' effect. Static elements or elements that move slowly with respect to the image frame are expected to benefit from this approach. The results of this application will be presented on the next section.

\section{Approach II: Probabilistic Approach}\label{sec:attentionModule}
Apart from a bare combination of the segmentation maps obtained at different time frames (section \ref{sec:imageBuffer}), there is also the possibility of making a more educated combination by modifying the probability map of each frame prior to the combination, resulting in an augmented segmentation. This approach is inspired by the transition probabilities in Markov Models that model the probability of an event transitioning between different possible states (section \ref{sec:MM}).\par
The intuition for this approach comes from the following line of thought: 'given that machine learning models are able to classify by calculating a confidence score (the logits) over a set of predefined labels, can those values be extended overtime (video inference) and used to influence the classification of consecutive frames?'. In other words, is it possible to establish a causal relation between past and present frame pixel-classification?\par
There are two problems that need to be addressed in order to create a temporal relation on the semantic image segmentation classification. First, the logit-map of each frame needs to be obtained and analyzed. Second, a relation between consecutive frames has to be defined.\par

\subsection{Semantic image segmentation - Scoremap}
\begin{figure}[ht]
    \centering
    \includegraphics[scale=0.5]{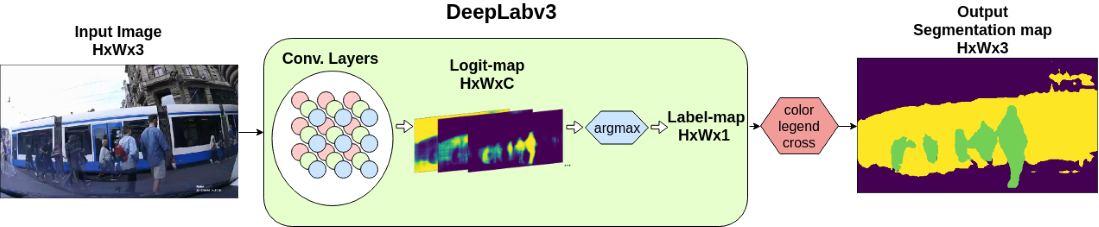}
    \caption{Graphical interpretation of the DeepLabv3 \citep{Chen2017RethinkingAC} semantic image segmentation tool-chain. The convolutional layers produce a HxWxC matrix (the logit-map) that is reduced by means of an argmax function into a HxWx1 matrix with the final labels of each pixel (the label-map). Finally an external function assigns a color value to each label.}
    \label{fig:scoremap-diagram}
\end{figure}

In semantic segmentation, the logit-map is a HxWxC matrix that contains, for each pixel in the image (HxW) a vector of real numbers that represents the likelihood of belonging to each one of the predefined categories (C), the logits. The logit-map is located one step before the final assignation of labels (figure \ref{fig:scoremap-diagram}) and is formed by real numbers without any apparent bounds.\par

In order to establish a relation between consecutive frames, it is necessary to transform the logits of each pixel to a common scale for comparison. This can be done by means of a softmax function (equation \ref{eq:softmax}) that transfer the raw prediction values of the neural network (the logits) into probabilities (logit-map into a probability-map).\par
\begin{equation}
    S(C_i)=\frac{e^{C_i}}{\sum_j e^{C_j}}
    \label{eq:softmax}
\end{equation}
where $C_i$ refers to each element in the logits vector C.\par
Once that the probabilities have been determined, it is possible to establish a relation between consecutive frames.

\subsection{Modeling videos as a weighted sum of frames} \label{sec:weightedsum}
% Inspired by the transition probabilities of Markov Models (section \ref{sec:MM})
The Attention Module models the classification probability of the present as a weighted sum of consecutive frames (equation \ref{eq:weightedsummation}). The number of frames in consideration for the sum is four (three past frames plus the present frame), this is a conservative choice based on our observations on the test videos.\par

\begin{equation}
    \textit{aug}=I_0\cdot p(t)+I_1\cdot p(t-1)+I_2\cdot p(t-2)+I_3\cdot p(t-3) 
    \label{eq:weightedsummation}
\end{equation}

\begin{figure}[ht]
    \centering
    \includegraphics[scale = 0.48]{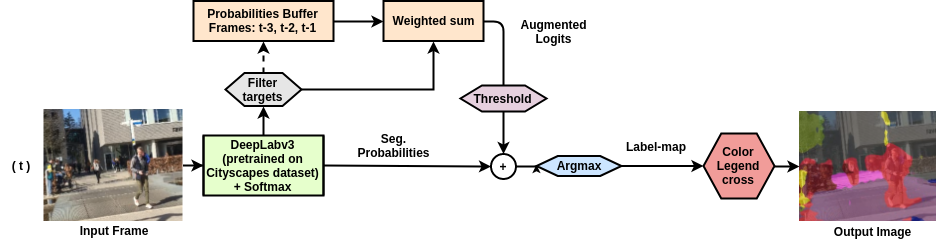}
    \caption{Block diagram of the Attention Module approach (section \ref{sec:attentionModule}). This block diagram depicts how the classification probabilities of 4 consecutive frames are combined into one.}
    \label{fig:AM2}
\end{figure}

$\textit{aug}$ is the resultant augmented logits (figure \ref{fig:AM2}), $I_n$ are the weights of frame $(t-n)$ and the probabilities $p(t-n)$ are extracted from the baseline segmentation model (section \ref{sec:deeplabv3}) after applying the transformation of the softmax function.

\subsubsubsection{Weights estimation}
In theory, it should be possible to find the optimal weight values that shape equation \ref{eq:weightedsummation} resulting in the probabilities with the smallest discrepancy with respect to the true segmentation. This is an optimization problem that requires the definition of a loss function and data-set with ground-truth annotations. In classification tasks the most common loss function is the cross-entropy function or the negative log likelihood\citep{cross-entropy1} (equation \ref{eq:cross-entropy})
\begin{equation}
        L = - \sum_{c=1}^M y_{o,c} log(p_{o,c}) \label{eq:cross-entropy}
\end{equation}
where $M$ is the number of classes; $y_{o,c}$ is a binary indicator that the class label $c$ is the correct classification for the observation $o$; and $p_{o,c}$ is the predicted probability of observation $o$ being of class $c$ \citep{cross-entropy}.\par
As opposed to the more common quadratic loss function, the cross-entropy loss function does not suffer slow learning due to the computation of small gradients \citep{cross-entropy1}. The gradients computed on the cross-entropy loss function increase when the prediction moves further from the target value \citep{cross-entropy1}. \par
Besides the definition of a loss function, it is also necessary to count with a ground-truth annotation data-set. However, as mentioned in section \ref{sec:testingandev}, no public multi-label semantic video segmentation data-set was found and developing one was not feasible due to the lack of resources. As an alternative, in order to find the 'best' weight values, the equation \ref{eq:weightedsummation} was tested using different parameter combinations on some of the videos listed in section \ref{sec:testingandev}. The value of the final weights  can be found in table \ref{tab:influenceParams}.\par
\begin{table}[ht]
    \centering
    \begin{tabular}{|c|c|c|c|c|}
        \hline
        $I_0$ & $I_1$ & $I_2$ & $I_3$ & $T$\\
        \hline
        $4$ & $3$ & $2$ & $1$ & $1$\\
        \hline
    \end{tabular}
    \caption{Weight parameters for equation \ref{eq:weightedsummation}}
    \label{tab:influenceParams}
\end{table}
As a result, using positive weights (bigger than 1) increases the sensitivity of the per-pixel classification, which added to the combination of successive frames result in a more consistent temporal segmentation (section \ref{sec:quantitative}). However, the increase in the sensitivity provokes a higher number of false positive classifications in the form of noise. The amount of false positive labels can be alleviated by setting up a threshold that pushes the augmented logits to zero if a minimum value is not reached (figure \ref{fig:AM2}). Similar to the Image Buffer, this approach is only applied to a certain category targets as a proof that it can be used to leverage the classification of any category labels. \par

\subsection{Experiments}
The experiments are divided in two parts: a quantitative evaluation that measures the performance of each method using different metrics and a qualitative evaluation that is done by interpretation of the segmentation results.\par

\section{Results}
This section collects and analyzes the results for all the experiments listed in section \ref{sec:testingandev}. It also critically analyzes the followed methodology highlighting the best case scenarios and revealing the possible failure modes.\par

\subsection{Quantitative results - Davis data-set}\label{sec:quantitative}
This section evaluates the performance of both suggested methods (Image Buffer and Attention Module), comparing them with the baseline segmentation (using DeepLabv3). Since the suggested methods are designed to \textit{reduce the number of false negative by leveraging the temporal information encoded in image sequences-videos}, this section looks at the segmentation performance measuring and plotting the \textit{Area overtime} (AOT) and \textit{IOU overtime} (IOU-OT). In addition, the frame to frame variation of both metrics will be calculated and plot side by side, expecting to find less variations on the approaches that handle past time information.\par
In order to quantify and compare the performance of each method, it is necessary to count with ground-truth annotations. However, it was not possible to find a data-set that covered this need and it was not feasible to create a custom one given the time-line. For this reason, the DAVIS data-set \citep{Perazzi2016} is used as an approximation with the limitation it can only evaluate one label at a time. The list of categories covered for this objective can be found in section \ref{sec:testingandev}.\par
\begin{figure}[ht]
    \centering
    \includegraphics[scale=0.5]{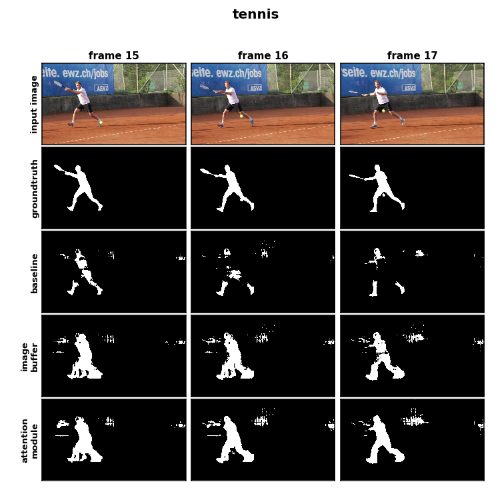}
    \caption{Figure comparison of the segmentation of 3 consecutive frames and its corresponding groundtruth annotation. It can be observed how both the Image Buffer and the Attention Module achieve a better segmentation of the 'person' category although there is also an increase in the amount of false positive classifications. Sequence source: DAVIS dataset \citep{Perazzi2016} - 'tennis' category.}
    \label{fig:tenniscomparisonch5}
    \includegraphics[scale=0.55]{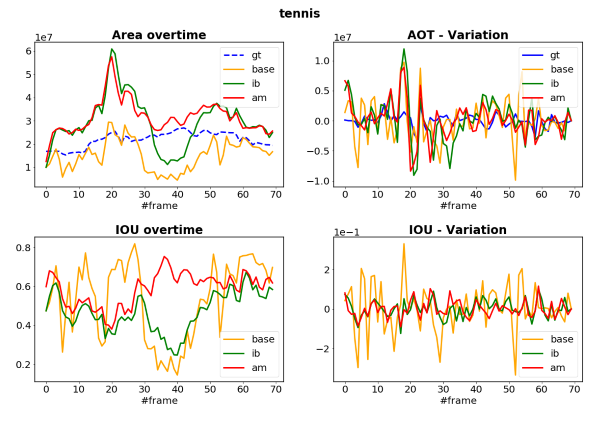}
    \caption{Graph evaluation of the 'tennis' category (DAVIS dataset \citep{Perazzi2016}) using the metrics defined in section \ref{sec:testingandev}. The upper graphs shows how the AOT of both suggested approaches is higher than the baseline model, although they tend to have a lower variation. The lower graphs show how both of the suggested approaches tend to be less accurate than the baseline segmentation, this can be reasoned by the increased production of false positive classifications. In addition, the IOU variation for both the IB and AT is lower than for the baseline.}
    \label{fig:tennismetricsch5}
\end{figure}

 
Figures \ref{fig:tenniscomparisonch5} and \ref{fig:tennismetricsch5} give an example of the segmentation evaluation for the 'tennis' category of the DAVIS data-set \citep{Perazzi2016}. Figure \ref{fig:tenniscomparisonch5} shows an obvious result on how leveraging neighbouring frames benefits the final segmentation output. In this first figure in can be observed how the baseline segmentation is surpassed by both of the suggested methods, achieving a smoother shape with the Attention module approach. The main difference between both of the suggested methods is that the Image buffer directly uses the segmentation produced by the baseline model, while the Attention module by modifying the probability map of the segmentation is able to produce new segmentation labels that might benefit the final segmentation. In figure \ref{fig:tenniscomparisonch5} it can also be observed how both of the suggested methods increase the number of false positive classifications, although it was intended to limit this number the study of its reduction is left as future work.\par
Figure \ref{fig:tennismetricsch5} contains the graphs that measure the AOT, the IOU-OT and its variations for the 'tennis' category of the DAVIS data-set. The AOT is calculated to compare the differences in the production of labels of each method. As expected, it can be observed how the area detection for both of the suggested methods is superior than the baseline detection and the ground-truth annotations. The reason for this is the temporal combination and the increase in the production of false positive classifications. The IOU-OT indicates the accuracy of each method over time, expecting values closer to 1 for the best case scenarios. The results show how both the Image Buffer and the Attention Module generally reproduce a lower accuracy than the baseline method, which can be explained due to the increase of false positive classifications. Lastly, the variation of each metric is computed to asses the reliability of each method (a reliable application is expected to give consistent results overtime). For a perfect segmentation, the AOT variation should be similar to the ground-truth annotations and the IOU-OT variation equal to zero.\par
The dispersion of the variation of each metric is calculated by computing the standard deviation of the sequence; tables \ref{tab:AOT Variation STD} and \ref{tab:IOU Variation STD} gather these results. The variation is computed to measure the frame-to-frame consistency of the prediction, a segmentation that fully captures the dynamics of the sequence (that is consistent overtime) would result in a AOT Variation standard deviation equal to the ground-truth's one and a IOU-OT Variation standard deviation equal to zero.\par
\begin{table}[ht]
    \centering
    \begin{tabular}{|c|c|c|c|c|}
    \hline
    \multicolumn{1}{|c|}{\multirow{2}{*}{}} & \multicolumn{4}{c|}{AOT Variation STD}\\ \cline{2-5} 
    \multicolumn{1}{|c|}{} & \multicolumn{1}{c|}{\textbf{Groundtruth}} & \multicolumn{1}{c|}{\textbf{\begin{tabular}[c]{@{}c@{}}DeepLabv3\\ (Cityscapes)\end{tabular}}} & \multicolumn{1}{c|}{\textbf{\begin{tabular}[c]{@{}c@{}}Image\\ Buffer\end{tabular}}} & \multicolumn{1}{c|}{\textbf{\begin{tabular}[c]{@{}c@{}}Attention\\ Module\end{tabular}}} \\
    \hline
        \textbf{Breakdance-flare} & 2297873 & 4771063 & 3649613 & \textbf{3347300}\\
    \hline
        \textbf{Bus} &1441019 & 4600729 & 3042283 & \textbf{2936702}\\
    \hline
        \textbf{Car-shadow} & 285037 & 1388070 & \textbf{1025686} & 1219912\\
    \hline
        \textbf{Car-turn} & 1128675 & 1767648 & 1221533 & \textbf{1212808}\\
    \hline
        \textbf{Hike} & 342290 & 1992053 & \textbf{1044184} & 1268832\\
    \hline
        \textbf{Lucia} & 443833 & 1430598 & \textbf{991134} & 1453644\\
    \hline
        \textbf{Rollerblade} & 666624 & 3548873 & 3240213 & \textbf{2480139}\\
    \hline
        \textbf{Swing} & 1663433 & \textbf{4711279} & 4975625 & 4908373\\
    \hline
        \textbf{Tennis} & 1073167 & 3690312 & 3523527 & \textbf{2904218}\\
    \hline
    \end{tabular}
    \caption{AOT Variation STD for some of the DAVIS data-set \citep{Perazzi2016} categories evaluated using different semantic segmentation approaches (section  \ref{sec:testingandev}). Although the three approaches under study are far from achieving a Variation close to the groundtruth, the consistency of both the Image Buffer and the Attention Module outperforms the bare DeepLabv3 implementation.}
    \label{tab:AOT Variation STD}
\end{table}
\begin{table}[ht]
    \centering
    \begin{tabular}{|c|c|c|c|}
    \hline
    \multicolumn{1}{|c|}{\multirow{2}{*}{}} & \multicolumn{3}{c|}{\textbf{IOU-OT Variation STD (\%)}}\\ \cline{2-4} 
    \multicolumn{1}{|c|}{} & \multicolumn{1}{c|}{\textbf{\begin{tabular}[c]{@{}c@{}}DeepLabv3\\ (Cityscapes)\end{tabular}}} & \multicolumn{1}{c|}{\textbf{\begin{tabular}[c]{@{}c@{}}Image\\ Buffer\end{tabular}}} & \multicolumn{1}{c|}{\textbf{\begin{tabular}[c]{@{}c@{}}Attention\\ Module\end{tabular}}} \\ 
    \hline
        \textbf{Breakdance-flare} & 8.65 & 4.64 & \textbf{4.11}\\
    \hline
        \textbf{Bus} & 4.58 & \textbf{2.82} & 2.92\\
    \hline
        \textbf{Car-shadow} & 2.97 & \textbf{1.75} & 1.81\\
    \hline
        \textbf{Car-turn} & 1.53 & 0.93 & \textbf{0.89}\\
    \hline
        \textbf{Hike} & 4.68 & 1.57 & \textbf{1.49}\\
    \hline
        \textbf{Lucia} & 2.85 & \textbf{1.95} & 2.13\\
    \hline
        \textbf{Rollerblade} & 14.67 & 7.25 & \textbf{5.89}\\
    \hline
        \textbf{Swing} & 5.56 & 3.69 & \textbf{3.78}\\
    \hline
        \textbf{Tennis} & 12.12 & 4.43 & \textbf{4.20}\\
    \hline
    \end{tabular}
    \caption{IOU-OT Variation STD for some of the DAVIS data-set \citep{Perazzi2016} categories evaluated using different semantic segmentation approaches (section  \ref{sec:testingandev}). A robust segmentation would be translated in having a low STD, which means that the accuracy does not fluctuate over the sequence. It can be observed how the lowest values are obtained by the two approaches that exploit temporal information, the Image Buffer and the Attention Module.}
    \label{tab:IOU Variation STD}
\end{table}

Although the results shown in this section are a simplification (single-label classification) of what it might be present in a real driving scenario, these results can serve as a guideline on how the combination of frames affects to the final segmentation. Next section presents the results for a series of multi-label classification experiments.

\subsection{Qualitative results - Video evaluation}\label{sec:qualitative}
In this subsection, the testing is extended to cover the multi-label detection. Due to the lack of groundtruth annotations, these experiments will be evaluated by the visual-quality of its results.\par
\begin{figure}[ht]
    \centering
    \includegraphics[scale=0.5]{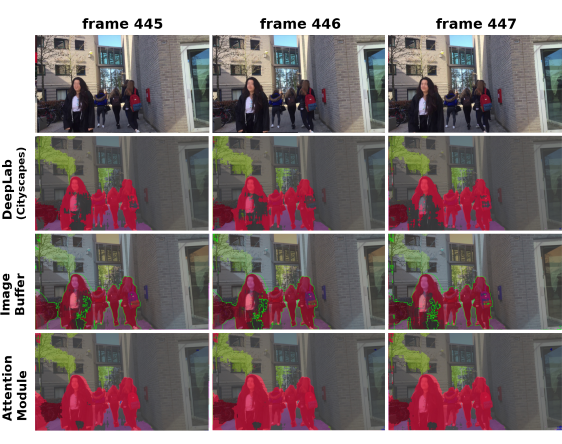}
    \caption{Figure comparison of the segmentation of three consecutive frames for a video sequence recorded outdoors in the presence of multiple target categories. It can be observed how the baseline segmentation (DeepLab) is able to detect most of the subjects and how the suggested extensions help completing the segmentation.}
    \label{fig:citadelcomparison}
\end{figure}

Figure \ref{fig:citadelcomparison} shows the segmentation performance of the baseline model (DeepLabv3) and its extension with the image buffer approach (detected area is contoured in red or blue) and the attention module approach. In particular, the extensions only leverage a set of predefined categories over the rest: \textbf{person, rider, bicycle, car and bus}. This is done as a proof of concept as well as due to hardware limitations, but it can be adapted to hold any other target categories.

\subsubsection{Citadel - University of Twente}
This video was taken on a normal day between lectures at the University of Twente. It was chosen to evaluate the performance of these methods in an environment rich on features with different targets present at the same time. Figure shows how both of the extensions help completing the partial segmentation of some targets.

Next section elaborates on the limitations of each method and makes a conclusion based on the results.

\section{Discussion}\label{sec:limitations}
% \subsection{Limitations}
The results show how the Image Buffer and Attention Module approaches can support the segmentation of a set of predefined target categories, this is due to the combination of information from neighbouring frames into the present segmentation.\par
Both of the extensions combine the segmentation of 3 past frames with the present frame. This is the result of the study performed and can only be applied to study-cases where the target object location does not differ considerably for 4 consecutive frames. In other words, it only applies to cases with a high recording-speed/target-relative-movement (w.r.t the camera) ratio. A low ratio will result on a 'ghosting effect' on the segmentation and a diffuse boundary definition of the targets.\par
The segmentation model also affects the final performance of both of the suggested extensions. Both of the extensions base their results on an initial baseline segmentation that is lastly modified. In the case of the Image Buffer this limitation is not very extreme and the final accuracy directly depend on the efficiency of the baseline model. In the other hand, the Attention Module modifies the probability-map of each frame by its multiplication with different weights (positive, greater than 1), this helps increasing the sensitivity of the detection of the target categories but in turn may produce false positive classifications.\par
Another limitation of the Attention Module approach is that its parameters were calculated heuristically, and the segmentation result might be sub-optimal. This was inevitable due to the lack of a sequential ground-truth data-set that could be used to find the optimal values of the weights and the threshold value (table \ref{tab:influenceParams}).\par
The chosen data-set for training the baseline model also affects the final performance of the segmentation. It determines the final accuracy of the segmentation and sets some limitations on the resolution of the images and the covered scenarios. Cityscapes data-set, covers driving scenarios in urban areas which might be the reason why the baseline segmentation offers a low performance in figures, due to the angle of the camera. Besides that, Cityscapes gathered samples during spring and summer seasons, so it does not include adverse weather conditions nor poor illumination images.\par

Nevertheless, both of the extensions have proved how leveraging sequential information for the application of real-time predictions is beneficial for the consistency of the results. When applied to autonomous navigation, creating an image segmentation training data-set that covers all the possible scenarios is very difficult and holding the information of past frames adds a protection layer against failure modes.\par

\section{Conclusions}
\label{ch:six}

The evaluation of the baseline performance, showed that the output segmentation over a sequence is irregular. This is a problem when the application relies on the frame segmentation as a main source of information. In order to attenuate these variations, both of the extension approaches (Image Buffer and Attention Module) combine the information of neighbouring frames to make a more educated segmentation of the present. Tables \ref{tab:AOT Variation STD} and \ref{tab:IOU Variation STD} show the effect of these extensions would successfully reduce the fluctuations on the segmentation overtime.\par
This study also showed some cases in which the baseline model was not able to detect a target in the field of view. Our new frameworks proved to reduce the number of false negative classifications. The Image Buffer approach sets a time lower bound on the segmentation providing a safety margin over sudden faulty modes. The Attention Module, due to the modifications on the probability map, is able to increase the sensitivity of the detection of a set of predefined targets while at the same time providing the advantages of the Image Buffer approach. Which can be very useful for the detection of obstacles in the case of autonomous navigation.\par

Both of the extensions (Image Buffer and Attention Module) require to store an array of dimensions: $MxHxWxN$. Where $M$ is the number of target categories, $H$ and $W$ the dimensions of the image source and $N$ the number of frames to be combined. This can be computationally very costly to be evaluated in a real-time implementation and requires a well selected hardware framework.\par
We can also suggest to use this methodology during the training phase of the segmentation model. The final user can create artificial annotations for the categories that appear to be more challenging for the baseline model and use them as ground-truth annotations for further training. This would expect to improve the segmentation over those target categories and get rid of the computational constraint imposed by having to store an array for the segmentation augmentation.\par    

As a future work, we believe that different ways using the suggested frameworks could be studied. Instead of overlapping the segmentation results of the baseline model or making a weight sum relation between consecutive frames, a more elaborated relation that uses probabilistic relations can be defined.\par
The combination of sequential information and enhancement of the probability-map results in an increase of false positive classifications in the form of segmentation noise. There are different ways that can help alleviate this problem, while at the same time raising the overall accuracy of the segmentation. Some of them are:
\begin{itemize}
    \item Fine tuning the parameters that define the Attention Module. In such a way that, the values that do not reach a minimum after the probability augmentation are pushed to 0 and therefore will not be assigned the target label.
    \item Using different temporal modules with different sensitivity values. After the computation of the augmented probabilities, the results will be compared and only the classifications with the highest agreement will be output.
    \item Limiting the work-space of the temporal modules. The augmented segmentation can be applied only to a certain region of interest, such as just the driving-space. This can alleviate the production of false positive classifications and increase the performance of the algorithm.
    \item Combining this method with a 3D map estimation. After the calculation of the augmented segmentation map, it could be projected onto a 3D map of the environment, assigning labels to the 3D objects and discarding the labels that do not project into any plane (or very far away).
\end{itemize}
Another extension of this study could be to instead of basing the augmentation on only past frames, an estimated prediction of nearby future states (t+1, t+2 ...) could be calculated and added into the combination. At the same time, given a sequential fully annotated data-set could provide the optimal set of parameters that define the relations between adjacent frames.\par
On top of that, these extensions could be applied to create an artificial sequential data-set. Using the results of the baseline model with the extension on the temporal domain could be used to generate short annotated sequences. These sequences can then be used to train deep learning architectures that embed the temporal behavior on its structure. Deep learning models such as convLSTM \citep{Shi2015ConvolutionalLN}, GRFP \citep{Nilsson2016SemanticVS} or the Multiclass semantic video segmentation \citep{Liu2015MulticlassSV} can be live-trained with these temporal extensions, creating a self-supervised segmentation pipeline.\par
It could be interesting to study the performance of semantic image segmentation models and its temporal extension from the inference-speed point of view. Although this study was performed on the state-of-the-art semantic image segmentation model (highest in accuracy), other models designed to perform at much higher speed rates can influence the type of solution needed for the temporal analysis.\par
Finally, finding new applications where this method can be applied. One of the highlights of this study is that slow environments benefit from sequential information, helping to smooth and to complete partial segmentation results (figure \ref{fig:citadelcomparison}). Of course, applications with different types of sequential data can also be analyzed under the same suggested approaches. This indicates the potential impact of our new sequential frameworks in a large field of video and image analysis.\par

\bibliographystyle{unsrt}  
\bibliography{main}  %%% Remove comment to use the external .bib file (using bibtex).
\end{document}